# Transparent matte surfaces enabled by asymmetric diffusion of white light

Hongchen Chu†[1], Xiang Xiong†[1], Nicholas X. Fang[2], Feng Wu[1], Runqi Jia[1], Ruwen Peng*[1], Mu Wang*[1,3], Yun Lai*[1]

[1]National Laboratory of Solid State Microstructures, School of Physics, and Collaborative Innovation Center of Advanced Microstructures, Nanjing University, Nanjing, 210093, China

[2]Department of Mechanical Engineering, University of Hong Kong, Pokfulam Road, Hong Kong

[3]American Physical Society, 100 Motor Pkwy, Hauppauge, NY 11788, USA

† These authors contributed equally to this work.
*Corresponding authors: Yun Lai (laiyun@nju.edu.cn); Ruwen Peng (rwpeng@nju.edu.cn); Mu Wang (muwang@nju.edu.cn)

**Summary paragraph**

The traditional wisdom for achieving transparency is to minimize disordered scattering within and on the surface of materials, so as to avoid translucency. However, the lack of disordered scattering also deprives the possibility of achieving a matte surface, resulting in the specular reflection and glare on transparent materials as a severe light pollution issue. In this work, we propose a solution utilizing optical metasurfaces[1-2] to overcome this long-existing dilemma. Our approach leverages an asymmetric background in metasurface design to achieve highly asymmetric diffusion of white light, maximizing diffusion in reflection while minimizing it in transmission across the entire visible spectrum. Using industrial lithography, we have created macroscale transparent matte surfaces with both strong matte appearance and clear transparency, defying the conventional belief that these two optical features are incompatible. These surfaces provide a remarkable phenomenon of switching between transparent or matte appearances via the brightness contrast between the front and rear ambient lights. They also support a unique application in transparent displays and augmented reality, offering perfectly preserved clarity, wide viewing angles, full color, and one-sided displays capabilities. Our findings usher in a new era of optical materials where the desirable properties of both transparent and matte appearances can be seamlessly merged.

With the explosive growth of the application of transparent materials such as windows, automotive glass and screens in modern life, the widespread problem of specular reflection and glare on such devices (Extended Data Fig. 1) becomes a significant issue of light pollution[3], causing physical discomforts like temporary blindness and dizziness. Despite that matte appearance has been proven effective in eliminating the specular reflection and glare from the surfaces of most opaque materials such as walls, wood, and unpolished metal, it is long believed that this strategy cannot apply to transparent materials. The physical reason is that in classical optics the random scattering or diffusion of light induced by rough surfaces or random scatterers is always achieved in both reflection and transmission channels, as shown in Fig. 1a. And the diffusion in transmission causes the blurring of transmitted images, thereby destroying transparency and resulting in translucent materials like frosted glass. Recent advances in metasurfaces[4-10], i.e. two-dimensional arrays of tailored artificial microstructures, offer an interesting solution to this long-exisiting dilemma through a random-flip configuration[11]. However, this design was hindered by some severe limitations such as microscale fabrication, inadequate bandwidth for the visible spectrum, and instability of diffusion efficiency due to dispersion. Additionally, most of the previous metasurfaces, including the random-flip one, did not take advantage of the asymmetric backgrounds in their designs, except for some special cases where high efficiency in transmission is pursued[12-16].

Here, we present a practical approach for realizing macroscale transparent matte surfaces (TMSs) for white light by utilizing extremely asymmetric diffusion across the entire visible spectrum. The asymmetric background of the metasurface plays a critical role in extending the bandwidth to encompass the whole visible range and stabilizing the degree of diffusion throughout the regime. These TMSs can be fabricated using industrial lithography, and their matte appearance and transmittance are adjustable, as demonstrated experimentally through several macro samples of 4 inches. Based on such macroscale TMSs, we further show a unique approach for transparent displays and augmented reality, offering features such as perfectly preserved clarity, wide viewing angles, full color, one-sided displays, and scalability to larger sizes.

The design principle of TMSs is described as follows. Physically, asymmetric light diffusion is achieved by applying phase shifts with random or uniform distributions in reflection or transmission, as illustrated in Fig. 1b. Inspired by previous research on coding metasurfaces[17-18] and random-flip metasurfaces[11] used for diffusing microwave or optical waves, we consider a metasurface composed of a randomly arranged set of two meta-atoms referred to as meta-atom I and meta-atom II. In this random binary optical system, the degree of diffusion could be characterized by $\xi = 1 - \frac{R_{spe}}{R_{tot}}$ in reflection or $\xi = 1 - \frac{T_{bal}}{T_{tot}}$ in transmission, respectively. Here $R_{spe}$ and $T_{bal}$ are the specular reflectance and ballistic transmittance, and $R_{tot}$ and $T_{tot}$ are the total reflectance and transmittance, i.e. the sums of the specular or ballistic part and the diffusive part in reflection or transmission, respectively. From the far-field radiation theory, we can calculate $\xi$ as a function of the phase difference $\Delta\varphi$ and amplitude ratio $\eta$ of the radiation from the two meta-atoms (see Supplementary Section 1). The averaged results calculated from 10 sets of random $100 \times 100$ configurations are shown in Fig. 1c. $\xi$ is almost unity when $\Delta\varphi = \pi$ and is near zero when $\Delta\varphi = 0$ under a broad range of $\eta$, i.e. $0.5 < \eta < 2$[11,17-19]. Therefore, in order to realize TMSs, the phase difference in reflection ($\Delta\varphi_r$) and transmission ($\Delta\varphi_t$) should be $\Delta\varphi_r \approx \pi$ and $\Delta\varphi_t \approx 0$ in the whole visible spectrum. Broadband $\Delta\varphi_t \approx 0$ can be achieved by using the reciprocity principle and local space inversion[11] or equivalent optical paths under certain circumstances. However, obtaining broadband $\Delta\varphi_r \approx \pi$ typically requires intricate structures with zero transmission[17-18], which are not suitable for our scenario that necessitates a manageable transmission to preserve transparency.

Utilizing an asymmetric background proves to be a viable solution for realizing a broadband phase difference of $\pi$ in reflection while maintaining control over the transmission. As shown in Fig. 1d, the optical meta-atoms consist of reflecting metal or dielectric patches embedded at varying depths beneath the dielectric surface. The optical path in the transmission is exactly the same through the two meta-atoms. Therefore we have $\Delta\varphi_t \approx 0$. On the other hand, the reflection coefficients of the two meta-atoms are approximately $r_1 = a + b_1$ and $r_2 = a + b_2$, respectively. Here $a$, $b_1$, and $b_2$ denote the contributions from the dielectric surface and the embedded patches in the reflection coefficient, respectively. For brevity, higher-order terms in reflection are ignored here because they are much smaller. $a$ is a real number determined by the refractive index of the dielectric substrate. By varying $d_1$ and $d_2$, it is possible to change the argument of $b_1$ and $b_2$ such that they are either in phase or out of phase to $a$, respectively, at a central frequency $f_0$, as shown in the left panel of Fig. 1e. The remarkable finding is that, as the frequency changes from $f_0$ to $f = f_0 + \Delta f$, the reflection coefficients $r_1$ and $r_2$ can remain nearly opposite to each other over a broad spectrum, as demonstrated by the pink shadow region in the right panel of Fig. 1e. Such a broadband $\pi$ phase difference attributes to the phase corrections of $\gamma_1$ and $\gamma_2$ caused by $a$. The mathematical proof of this mechanism is shown in Supplementary Section 2. We note that if the dielectric layers above the reflective patches are absent, i.e., $a = 0$, then the phase difference of π can only be obtained at a single frequency because the phase shifts of $b_1$ and $b_2$ under frequency change $\Delta f$, i.e., α and β, are distinct.

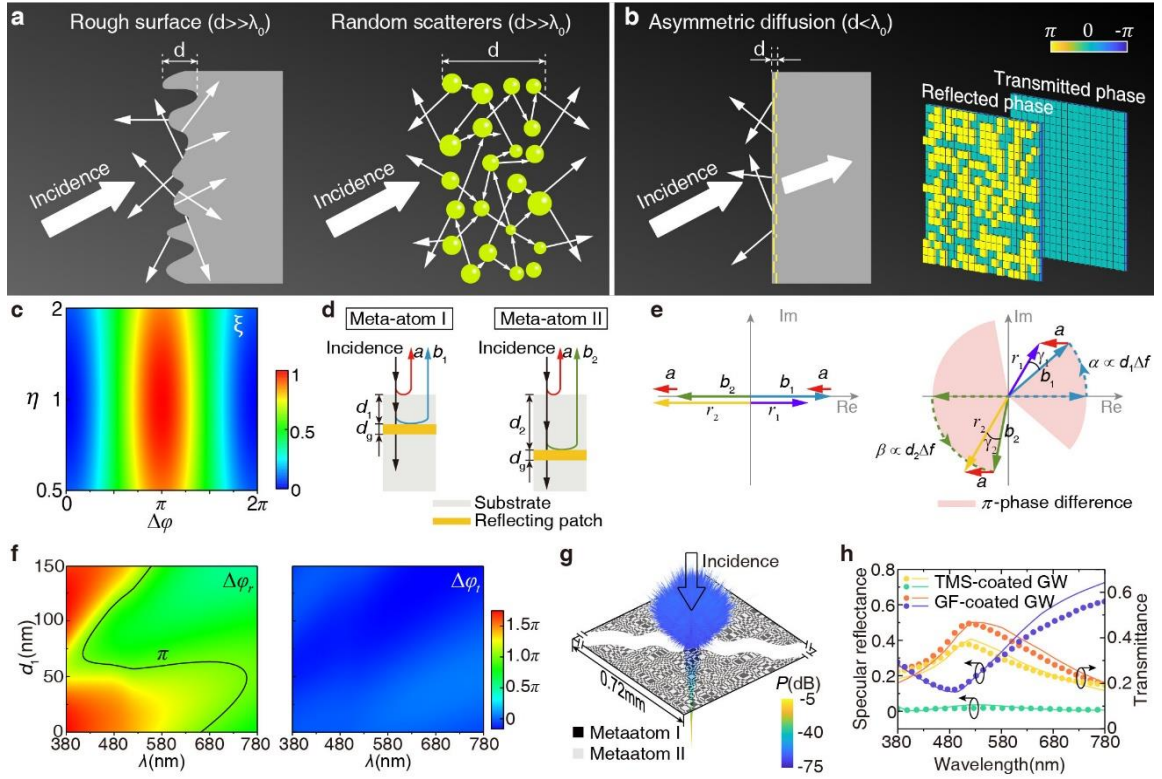

**Fig. 1 | TMSs enabled by extremely asymmetric diffusion of white light. a**, Traditional methods of achieving a matte surface, such as rough surfaces (left) and random scatterers (right), result in diffused light in both transmission and reflection channels. **b**, By introducing random phase shift in reflection and uniform phase shift in transmission, optical metasurfaces allow for asymmetric diffusion of white light, effectively merging matte appearance and transparency. **c,** The degree of diffusion as a function of the phase difference and amplitude ratio of the radiation from the two meta-atoms in 10 sets of random $100 \times 100$ configurations. **d**, The two meta-atoms are composed of reflecting patches (yellow) with a thickness $d_g$ located at depths $d_1$ and $d_2$ beneath the surface of the dielectric substrate (grey). **e,** The mechanism of achieving broadband near-π phase difference in the reflection coefficients $r_1 = a + b_1$ and $r_2 = a + b_2$. **f**, Calculated phase differences in reflection ($\Delta\varphi_r$) and transmission ($\Delta\varphi_t$) as functions of $d_1$ and wavelength $\lambda$ for meta-atoms with gold patches of $d_g = 28$ nm and $d_2 = d_1 + 90$ nm. $\Delta\varphi_r \approx \pi$ and $\Delta\varphi_t \approx 0$ are realized in the whole visible spectrum (380 nm − 780 nm) at $d_1 = 70$ nm. **g**, Calculated far-field radiation power (P) pattern at the wavelength of 413 nm for a unit of TMS (gray figure) composed of $800 \times 800$ random sequences of the two meta-atoms. **h**, Measured (symbols) and calculated (curves) specular reflectance and transmittance of

the gold TMS-coated GW of $d_g = 28$ nm, $d_1 = 70$ nm, and $d_2 = 160$ nm, and the GW coated with a gold film of the same thickness (28 nm).

Without loss of generality, we first consider meta-atoms consisting of gold patches of thickness $d_g = 28\ nm$ and choose the central wavelength to be $\lambda_0 = c/f_0 = 540$ nm, where $c$ is the speed of light. From the opposite phase condition between $b_1$ and $b_2$ at $f_0$, it is obtained that $d_2 = d_1 + 90$ nm. In Fig. 1f, we plot the calculated phase differences between the two meta-atoms in reflection ($\Delta\varphi_r$) and transmission ($\Delta\varphi_t$) as functions of $d_1$ and wavelength $\lambda$. It is observed that when $d_1 = 70$ nm, $\Delta\varphi_r \approx \pi$ and $\Delta\varphi_t \approx 0$ in the whole visible spectrum (380 nm − 780 nm). The design of the TMS unit involves arranging the two meta-atoms depicted in Fig. 1d in an $800 \times 800$ random pattern, measuring $0.72$ mm × $0.72$ mm in size. The metal patches are set to a lateral scale of 900 nm, which is small enough to diffuse light in the visible band and large enough to be fabricated using standard industrial lithography. The far-field radiation power pattern at a wavelength of 413 nm, calculated and depicted in Fig. 1g, shows a highly asymmetric (backward) optical diffusion. This is evident from the diversification of reflected light in many backward directions and the absence of diffusion in transmitted light. At other frequencies, the specular reflectance is approximately 1 %, mainly due to differences in the reflectance of the meta-atoms. We note that it is still significantly lower than that of the glass substrate (~7 %).

This TMS design has a significant advantage in that it can be easily mass-produced through industrial lithography, which is crucial for practical applications. We fabricate a macroscale TMS by repeating the TMS unit from Fig. 1g on a 4-inch glass wafer (GW). In the fabrication, we have chamfered the right angles of the metal structures to avoid potential singularity effects[20] and adopted an aligned double-exposure process to fabricate the two complementary layers of random metal patches, which in principle applies to any bilayer complementary structures[21]. More details of the TMS fabrication process are described in the Methods section. Experiments have been conducted to measure the transmittance and specular reflectance of a 28 nm thick gold TMS (with $d_g = 28$ nm, $d_1 = 70$ nm, $d_2 = 160$ nm) and a 28 nm thick smooth gold film coated glass wafer (GF-coated GW). The results align well with theoretical calculations, as shown in Fig. 1h. The measured average specular reflectance of the TMS-coated glass wafer is 1.3 %, which is significantly lower compared to 34 % for the GF-coated glass wafer and 7 % for the uncoated glass wafer. When the incidence is from the other side of the glass wafer, the broadband π phase difference is violated, resulting in an increase in the specular reflectance at specific frequencies, e.g. over 10 % at λ>677nm (as shown in Extended Data Fig. 2).

To evaluate the performance of macroscale TMSs, we experimented in a mini photographic studio. A circular hole in the studio was covered with a TMS-coated GW (28 nm, Au). A bottle of flowers was placed in front of the studio, as depicted in Fig. 2a. The zoomed-in image of the TMS, shown in the left panel of Fig. 2b, confirms its matte appearance, as the mirror image of the flowers is barely visible. The TMS is found to have a similar appearance to unpolished gold and does not change with the angle of observation, as demonstrated in Extended Data Fig. 3. For comparison, a commercial anti-glare film (AF) was used, and the photograph taken with the same conditions is shown in the left panel of Fig. 2c. The mirror images of the flowers are severely blurred, confirming the diffuse reflection. The right panel of Fig. 2b shows a clear photograph of the flowers, which was taken inside the photographic studio through the TMS-coated GW, confirming the perfect clarity of the TMS-coated GW as if viewed through an open hole. In contrast, a photograph taken through the AF under the same conditions is shown in the right panel of Fig. 2c, where the image of the flowers and bottle is significantly blurred as if viewed through frosted glass. This blurring effect would be apparent to the naked eye as long as the AF is not tightly attached to the observed objects, as demonstrated in Extended Data Fig. 4. The TMS-coated GW thus exhibits a unique combination of matte appearance and clear transparency, which were believed to be incompatible previously. We note that the transparent and matte appearance can be flexibly switched by controlling the brightness contrast between the front and rear ambient lights. This exceptional feature is also confirmed under outdoor conditions with natural light, as shown in Supplementary Video 1. We have also made a direct comparison of the functionalities of the GF-coated and TMS-coated GWs, and the AF, considering there are both objects placed in front of and behind them, as presented in Extended Data Fig. 5.

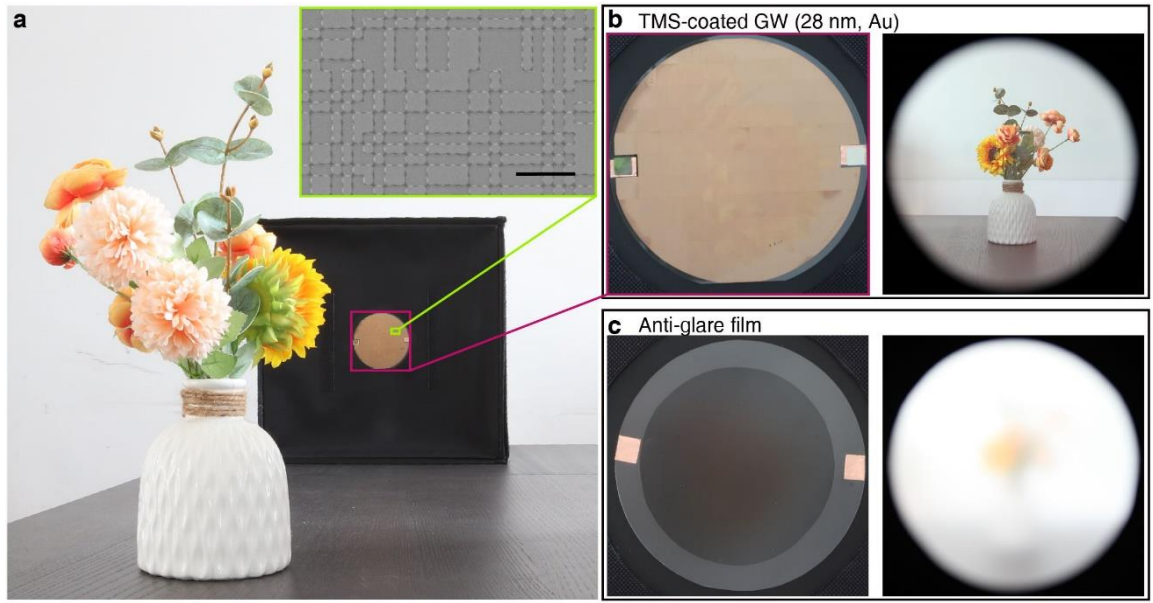

**Fig. 2 | Experimental demonstration of TMSs. a**, The experimental setup. The TMS-coated GW (28 nm, Au) was mounted on a hole of a mini photographic studio. Inset, SEM micrograph of part of the TMS. The scale bar represents 5 μm. **b**, (left panel) A zoom-in photograph of the TMS in **a**, where almost no mirror image is observed due to the matte appearance. (Right panel) A photograph taken inside the studio shows a clear image of the bottle of flowers viewed through the TMS-coated GW (28 nm, Au). **c**, (left panel) A photograph taken outside the studio of a commercial AF mounted on a hole, showcasing a severely blurred mirror image of the flowers. (Right panel) A photograph taken inside the studio through the AF displays a significantly blurred image of the flowers and bottle.

The appearance and transmittance of the TMS can be tailored by changing the material and thickness of the reflecting patches, respectively. In addition to the gold TMS ($d_g = 28$ nm) shown in Fig. 2a, we fabricated a titanium TMS ($d_g = 28$ nm) and gold TMSs with different thicknesses ($d_g = 17$ nm, $d_g = 35$ nm) on GWs using industrial lithography. The samples were placed over the hole and photographs were taken both outside and inside the studio, with the same camera parameters and environment as in Fig. 2b (Extended Data Fig. 6). From the photographs, it is seen that the TMS samples appear as matte surfaces in reflection, while also being clearly transparent. The appearance and transmittance are both customizable using different materials or structures in the TMS. For instance, the titanium TMS has the gray appearance of unpolished titanium, while the transmittance of the gold TMS decreases as the thickness of the reflecting patches increases. The measured and calculated transmission and reflection spectra of these TMSs are shown in Extended Data Fig. 7.

With both a matte appearance and clear transparency, TMSs offer a novel approach to transparent displays using image projection, as schematically demonstrated in Fig. 3a. The matte appearance of TMSs across the entire visible spectrum results in a wide-angle and full-color displays (Extended Data Fig. 8). In Fig. 3b, a colorful butterfly was projected onto a TMS-coated GW ($28 nm$, Au) with a real sunflower placed behind it. The butterfly and sunflower were both clearly visible. In contrast, when the TMS-coated GW was replaced with an AF, as shown in Fig. 3c, the image of the sunflower was blurred, and the brightness of the butterfly was much dimmer than on the TMS. An important application of transparent displays is augmented reality, as schematically shown in Fig. 3d. A webcam captures the image of an object behind the TMS, which is then analyzed by a pre-trained deep neural network (GoogLeNet). If the object is successfully identified, its name is projected, otherwise "unidentified" is displayed. As shown in Fig. 3e, a tennis ball was successfully recognized through the TMS-coated GW ($28 nm$, Au), and its name was displayed. In contrast, the blurring effect of the AF prevented such recognition, as shown in Fig. 3f (see additional examples in Supplementary Videos 2 and 3).

The extremely asymmetric light diffusion of TMSs also enables a one-sided displays. When the transmittance is not extremely low, the displayed content is usually observable from both sides of the projection screen due to the light diffusion in both reflection and transmission[22-23]. However, the asymmetric diffusion of TMSs significantly reduces diffusion in transmission, leading to high contrast in the brightness of the displays on the two sides. This was verified in the experimental setup in Fig. 3g. An image of a row of colored pencils was projected onto the TMS or AF, and the photographs were taken from both sides under the same camera parameters and combined in Figs. 3h and 3i, for the TMS-coated GW ($28 nm$, Au) and AF, respectively. Clearly, the displayed pencils on the TMS were much brighter on the side of the projection (front view), indicating that light diffusion was mainly confined to reflection. In contrast, the displayed pencils on the AF had comparable brightness on both sides and are markedly dimmer than that on the TMS. This unique one-sided displays helps protect the privacy of displays and enables the double-sided displays without crosstalk (see Extended Data Figure 9).

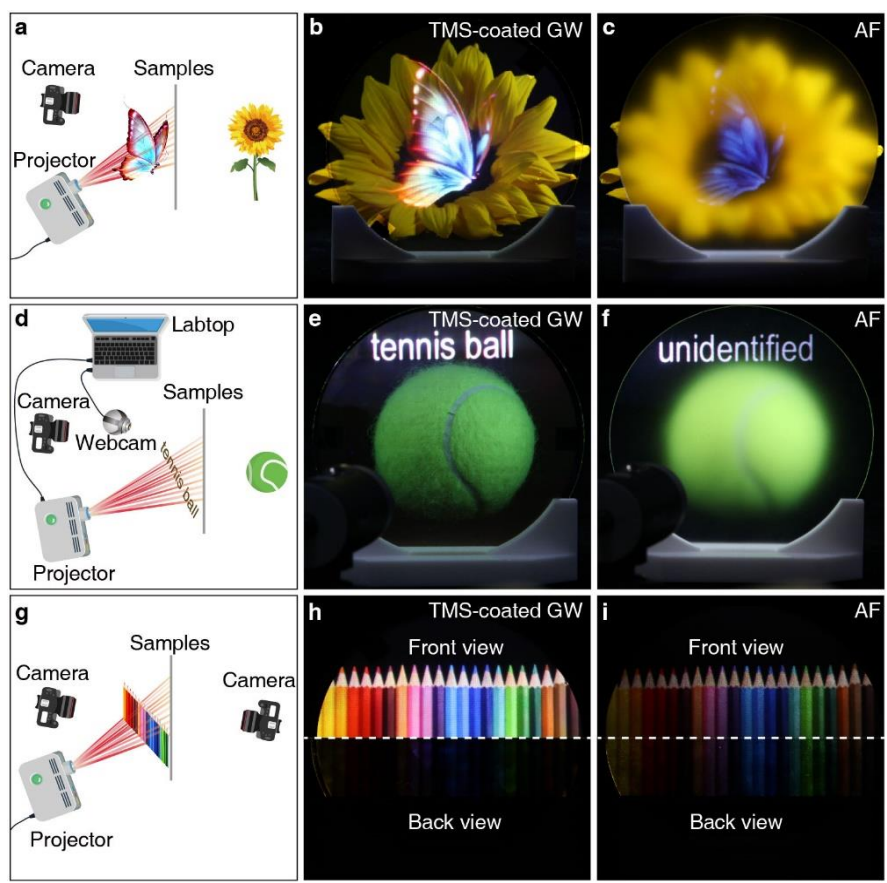

**Fig. 3 | Transparent displays and augmented reality by TMS. a**, Schematic diagram of the transparent displays based on TMS. The image of a butterfly was projected onto the TMS-coated GW or AF with a real sunflower behind it. **b,c**, Photographs taken under the same camera parameters for the TMS-coated GW (**b**) and AF (**c**). **d**, Schematic diagram of an augmented reality application. A webcam captures the image of a tennis ball placed behind the TMS-coated GW/AF, which is processed by machine vision technologies for recognition. **e,f**, Photographs taken under the same camera parameters for the TMS-coated GW (**e**) and AF (**f**). **g**, Schematic diagram for verifying full-color and one-sided displays. Photographs were taken from both sides of the TMS-coated GW or AF under the same camera parameters, with the image of a row of colored pencils projected from the front side. **h,i**, Combined photographs for the TMS-coated GW (**h**) and AF (**i**).

The reflecting patches in the TMSs can be made of metal, such as gold or titanium, but this principle can be easily extended to dielectric patches[24-27], resulting in a minimal loss. Nevertheless, we note that metal-based TMSs offer a remarkable advantage, i.e. good insulation for infrared thermal energy, which makes them useful for heat insulation in windows and vehicle glass (see Extended Data Fig. 10), similar to low-E glass[28]. Apparently, TMSs can avoid the increase of glare in traditional low-E glass coatings, due to their exceptional matte appearance.

The process of light diffusion[17-18] by metasurfaces is essentially the opposite functionality of cloaking a corrugated surface[29]. Compared with diffusion by random bulk media[30-34], this metasurface approach can avoid the phenomena of wave localization[30] and large absorption[31] due to subwavelength thickness. Recently, diffusive metasurfaces have provided unprecedented dynamic visual effects through spectrally and angularly shaping the reflected light[35-36]. While a fundamental difference here is that both the transmission and reflection channels are manipulated independently, rendering it possible to maintain the clarity of transparency while achieving a matte appearance. The concept of disorder engineering, which has had a significant impact in fields such as high-resolution imaging[37], optical spin-Hall effect[38], and high-capacity holograms[39], can be further applied to TMSs to refine and optimize the performance of diffusion. Additionally, the asymmetric background significantly reduces the complexity of meta-atom design for broadband phase manipulation[40-41] and facilitates large-scale fabrication through industrial lithography.

The concept of TMS breaks with the conventional wisdom that transparency can only be maintained by reducing disordered light scattering on a material's surface and within its body. Despite being comprised of random scatterers that exhibit strong diffuse reflection, TMSs retain clear transparency far beyond that of commercial AFs. This opens up new possibilities for applying anti-glare techniques to general transparent devices like windows and automotive glass, which was not possible with AFs because the transmitted image of distant scene would be severely blurred. Another amazing functionality is switching between transparent and matte appearance through the control of the brightness contrast between the front and rear ambient lights. This surface also offers a unique route for transparent displays and augmented reality with undamaged clarity in transparency, which is clearly beyond popular transparent display systems based on light-emitting devices and circuits, or random scatterers with untailored diffusion property.

## Methods

### Fabrication of the TMS sample

The TMS sample has been fabricated by multilayer aligned stepper photolithography on a 4-inch glass wafer with a thickness of 500 μm. First, lithography defined a randomly distributed patch pattern on the wafer with photoresist. Then, a metallic film with designed thickness is deposited by electron beam evaporation on the patterned wafer, covering both the areas with the photoresist and where the photoresist has been removed. Thereafter, the left photoresist was removed with solvent, leaving only the first layer of metallic patch pattern on the glass wafer. Then a 90-nm-thick silica layer is deposited by plasma-enhanced chemical vapor deposition (PECVD) as a spacer layer. The process continued with spin coating a photoresist layer, followed by a second lithography process. The second layer of metallic film was deposited, and a lift-off procedure was performed to remove the photoresist, leaving the second layer of the metallic patch pattern. Lastly, a 70-nm-thick silica layer is deposited to cover the entire sample area.

**Acknowledgements** The authors acknowledge financial support from the National Key R&D Program of China (Grants No. 2022YFA1404303 and No. 2020YFA0211300), the National Natural Science Foundation of China (Grants No. 12174188, No. 11974176, No. 12234010, No.11974177, and No.61975078), and N.F. acknowledges the startup funding from the Hong Kong Jockey Club Charitable Trust for Lab on Scalable and Sustainable Photonic Manufacturing at HKU.

**Data availability** All data needed to evaluate the conclusions in the paper are present in the paper and/or the Supplementary Information.

**Author contributions** H.C. and X.X. contributed equally to this work. Y.L., R.P., and M.W. organized and led the project. H.C. and Y. L. designed the metasurface and performed the theoretical analysis. X.X., R.P., and M.W. fabricated the transparent-matte-surface samples and performed the optical measurement. N.F. contributed to discussions of the data and analysis. F.W. and R.J. helped with the simulations. H.C., Y.L., M.W., R.P., and X.X. prepared the manuscripts.

**Competing interests** The authors declare no competing interests.

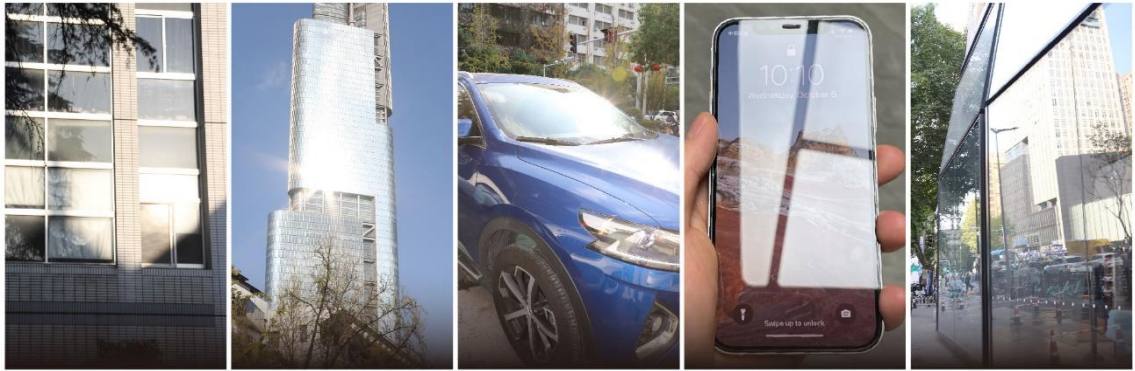

**Extended Data Fig. 1** | Specular glare and reflection on transparent devices result in a common type of light pollution in modern life.

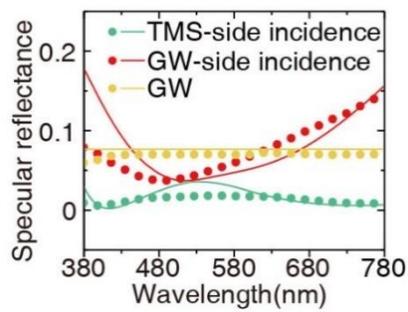

**Extended Data Fig. 2** | The measured (symbols) and calculated (curves) specular reflectance of a GW (yellow) and the TMS-coated GW (28 nm, Au) under the illumination of incidence from the TMS side (green) and GW side (red). As measured, the average specular reflectance of the TMS-coated GW under incidence from the GW side is 7.6 %.

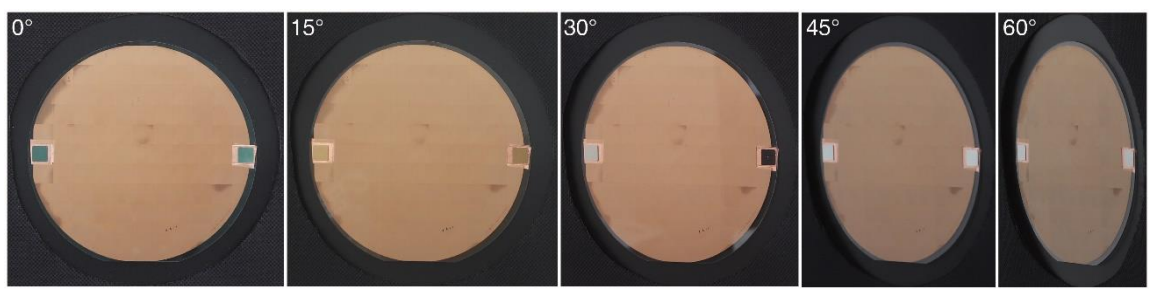

**Extended Data Fig. 3** | Photographs of a TMS-coated GW (28 nm, Au) taken outside the photographic studio at viewing angles of 0°, 15°, 30°, 45°, and 60°.

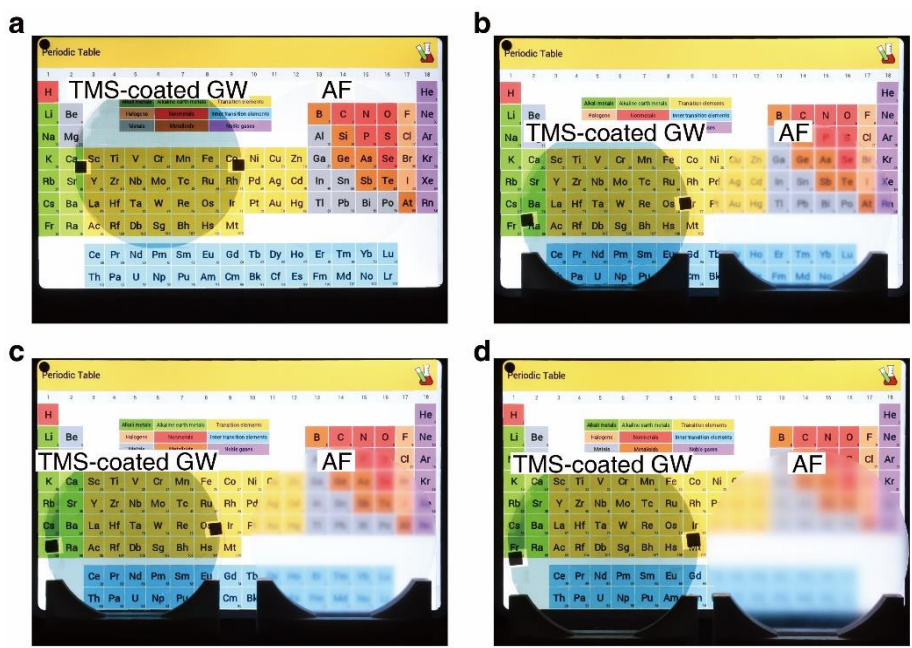

**Extended Data Fig. 4** | Photographs of a periodic table of elements behind a TMS-coated GW (28 nm, Au) and AF. The distance between the periodic table of elements and the TMS-coated GW/AF are separately 0 cm (**a**), 2 cm (**b**), 5 cm (**c**), and 10 cm (**d**).

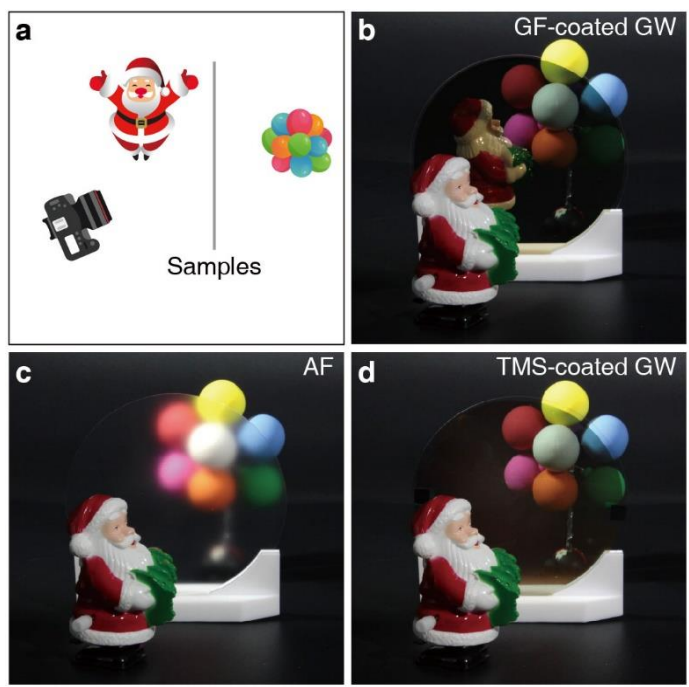

**Extended Data Fig. 5** | A direct comparison between the GF-coated GW, AF, and TMS-coated GW (28 nm, Au). **a,** A toy Santa and a bunch of colorful balloons are placed on each side of the sample. A camera on the toy Santa's side can take photographs of the sample and two toys. **b-d** Photographs of the two toys on each side of a GF-coated GW, an AF, and a TMS-coated GW. The reflected image of the Santa is clearly observed on the GF-coated GW (b), but disappears on the AF and TMS-coated GW (c, d). While the colorful balloons are clearly observed through the GF-coated GW (b) and TMS-coated GW (d) but is blurred through the AF (c).

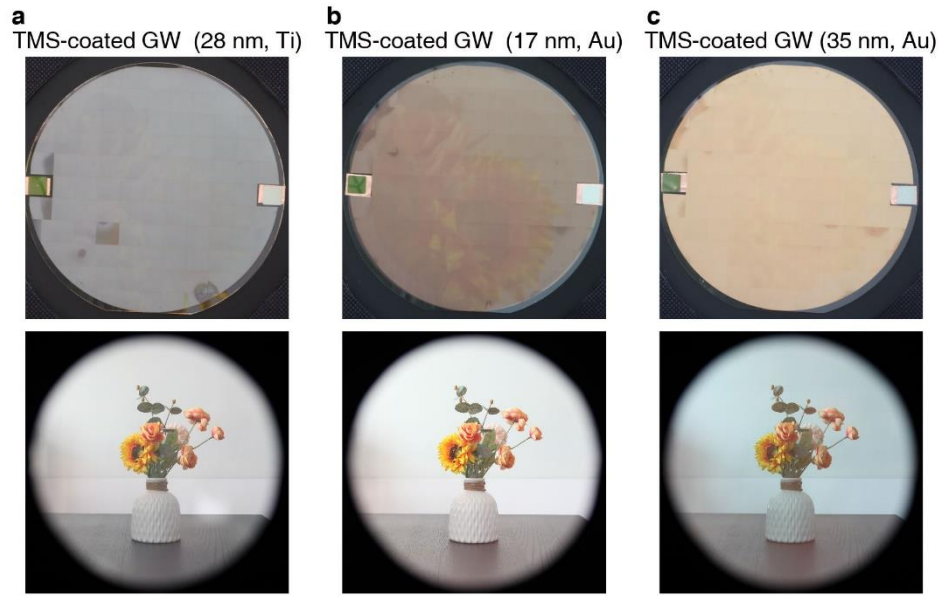

**Extended Data Fig. 6 |** Photographs taken outside (upper panels) and inside (lower panels) the studio of a titanium TMS-coated GW (28 nm, Ti) (**a**) and two gold TMS-coated GWs of different thicknesses (17 nm, Au) (**b**), and (35 nm, Au) (**c**) mounted on the hole, respectively.

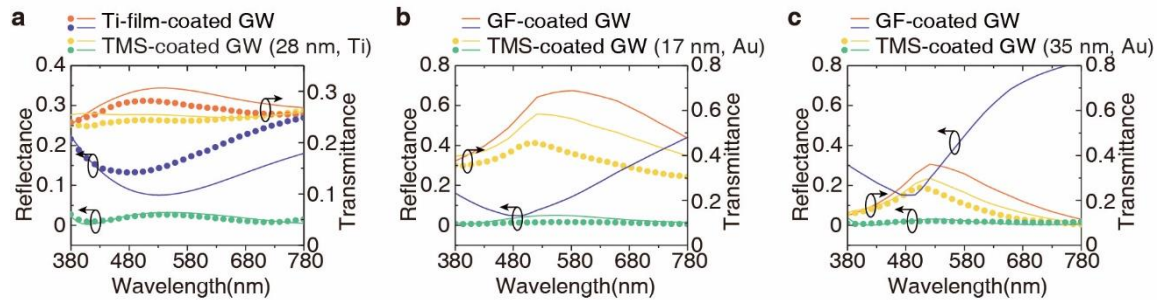

**Extended Data Fig. 7 |** Measured (symbols) and calculated (curves) specular reflectance and transmittance of the GW with a titanium TMS (28 nm) and two gold TMSs (17 nm, 35 nm) and the corresponding GW coated with titanium and gold films of the same thicknesses. As measured, the average specular reflectance of the three TMS-coated GW, are separately 1.7 % (28 nm, Ti), 1.3 % (17 nm, Au), and 1.7 % (35 nm, Au).

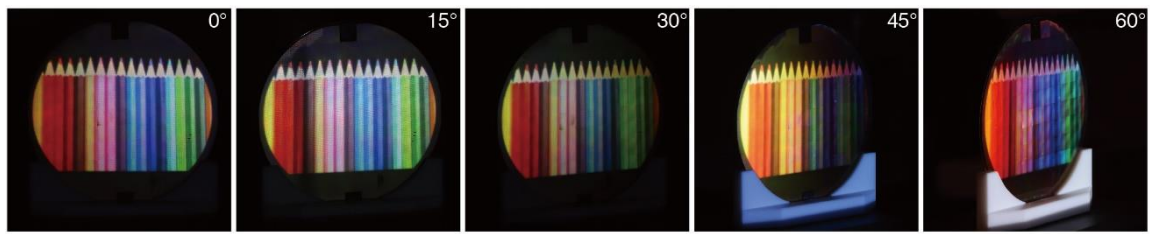

**Extended Data Fig. 8 |** Photographs of a TMS-coated GW (28 nm, Au) at viewing angles of 0°, 15°, 30°, 45°, and 60°, when the image of a row of colored pencils is projected on the TMS.

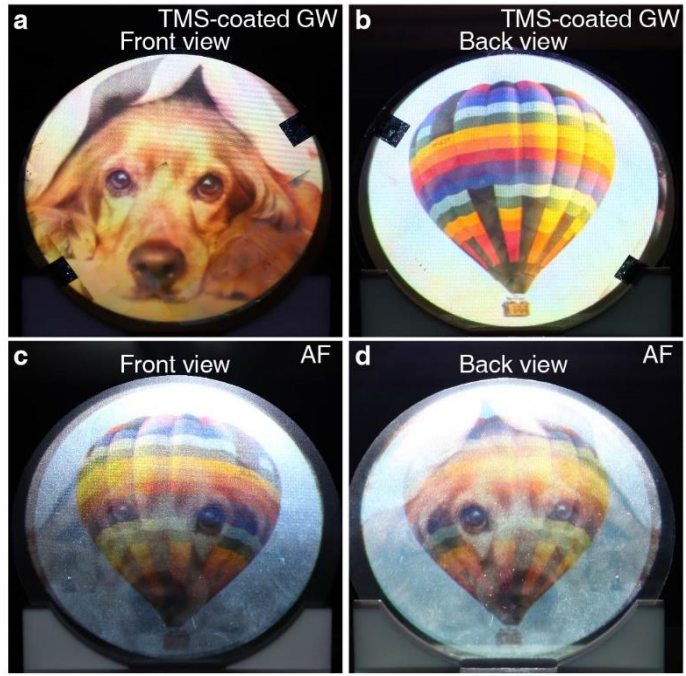

**Extended Data Fig. 9 | Experimental demonstration of the double-sided projection displays.** Images of a dog and a hot air balloon are simultaneously projected onto the front and back sides of the TMS-coated GW (28 nm, Au). **a,b,** Two independent images of a dog and a hot air balloon are separately observed on the front (**a**) and back (**b**) sides of the TMS-coated GW. **c,d,** The TMS-coated GW is replaced with an AF. Two blended images of the dog and hot air balloon are observed on both sides, indicating severe crosstalk between the displays on the front and back sides.

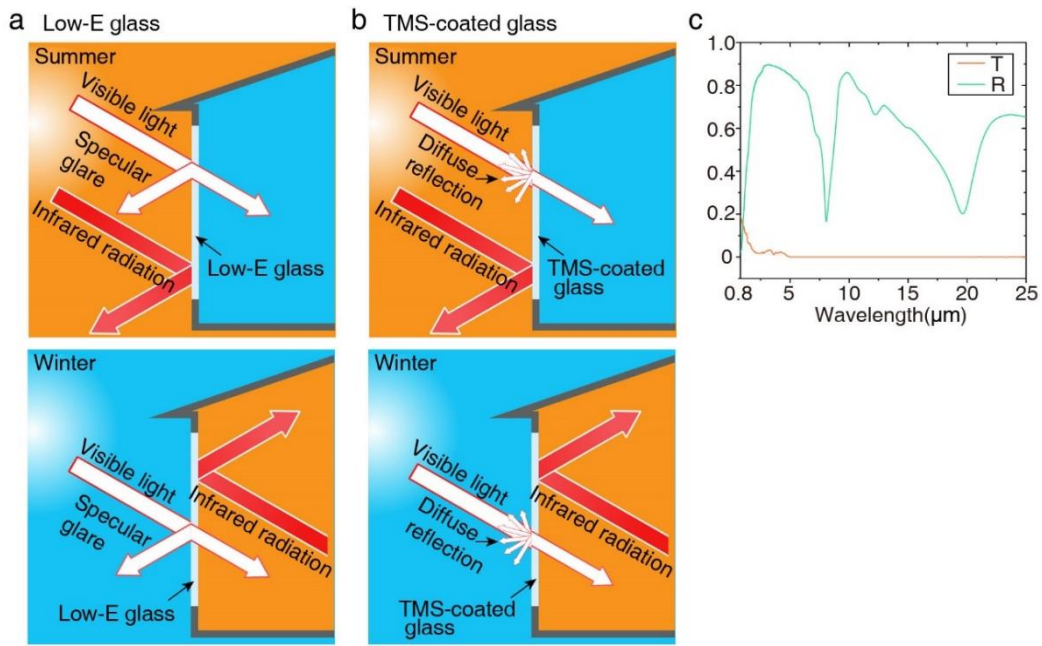

**Extended Data Fig. 10 | a,** Low emissivity (Low-E) glass introduces thin metal layers that block infrared radiation from the outdoors, resulting in a cooler room in the summer. Additionally, Low-E glass can keep the room warmer in the winter by preventing infrared radiation from escaping the room. However, the metal layers cause strong specular glare in the visible spectrum. **b,** The

TMS-coated glass blocks the infrared radiation similarly to the Low-E glass. In the visible spectrum, the TMS-coated glass is clearly transparent and can eliminate the strong specular glare via diffuse reflection. **c,** Measured transmittance (T) and specular reflectance (R) of the TMS-coated GW (28 nm, Au) in the infrared spectrum.

# Supplementary Information for

# Transparent matte surfaces enabled by asymmetric diffusion of white light

Hongchen Chu†[1], Xiang Xiong†[1], Nicholas X. Fang[2], Feng Wu[1], Runqi Jia[1], Ruwen Peng*[1], Mu Wang*[1,3], Yun Lai*[1]

[1]National Laboratory of Solid State Microstructures, School of Physics, and Collaborative Innovation Center of Advanced Microstructures, Nanjing University, Nanjing, 210093, China

[2] Department of Mechanical Engineering, University of Hong Kong, Pokfulam Road, Hong Kong

[3]American Physical Society, 100 Motor Pkwy, Hauppauge, NY 11788, USA

† These authors contributed equally to this work.
*Corresponding authors: Yun Lai (laiyun@nju.edu.cn); Ruwen Peng (rwpeng@nju.edu.cn); Mu Wang (muwang@nju.edu.cn)

**Section 1. Far-field radiation calculation and the optimization of the random sequence of the TMS**

For a TMS composed of a random configuration of two types of meta-atoms under the illumination of a normal incidence plane wave, each meta-atom can be treated as a secondary source with scattering amplitude $A$, scattering phase $\phi$, and pattern function $f_e(\theta,\varphi)$, where $\theta$ and $\varphi$ are the elevation and azimuth angles, respectively. Then the scattered far-field function of the TMS can be expressed as:

$$f(\theta,\varphi) = \sum_{m=1}^{M} A_m f_{e,m}(\theta,\varphi)\exp\{-i(\phi_m + k\sin(\theta)\cos(\varphi) x_m + k\sin(\theta)\sin(\varphi) y_m)\}, \quad (S1)$$

where $(x,y)$ is the coordinate of meta-atoms and subscript $m$ depicts the $m$th meta-atoms.

The specular reflectance $R_{spe}$ (ballistic transmittance $T_{bal}$) can be calculated by integrating the scattered far-field power function, i.e., $|f(\theta,\varphi)|^2$, near the direction of the specular reflection (ballistic transmittance), where the specific range of the solid angle for this integration depends on the width of the specular reflection (ballistic transmittance) lobe. Similarly, the total reflectance $R_{tot}$ (the total transmittance $T_{tot}$) can be obtained by integrating $|f(\theta,\varphi)|^2$ in the whole half space of reflection (transmission). Then the degree of diffusion could be calculated by $\xi = 1 - R_{spe}/R_{tot}$ in reflection or $\xi = 1 - T_{bal}/T_{tot}$ in transmission, respectively.

To generate an optimized 2-dimensional (2D) random pattern of the two types of meta-atoms in the TMS, firstly, we assume a random 1-dimensional (1D) sequence of meta-atoms with a lattice constant *D* comparable to the wavelength in vacuum, where the two kinds of meta-atoms are separately depicted by "0" and "1". The two meta-atoms are assumed to have identical $A$ and distinct $\phi$ with $\pi$ difference. The pattern function of each meta-atom, i.e., $f_e(\theta,\varphi)$ can be calculated through Eq. S1 by assuming that the meta-atom is composed of an array of identical deep-subwavelength scatterers, e.g., $20\times 20$ scatterers with a pattern function of $\cos(\theta)$. Then, the scattered far-field function of this 1D sequence can be calculated by Eq. S1. We employ the ergodic algorithm to optimize a 1D sequence with 20 meta-atoms by calculating the maximum value of $f(\theta,\varphi)$ for all $2^{20}$ cases and then selecting the top 40 sequences with the minimum $\text{Max}(f(\theta,\varphi))$, which corresponds to better diffuse scattering. Secondly, we randomly disrupt the order of the 40 sequences to form two 1D sequences of length 800. Thirdly, we apply the two optimized 1D sequences in $x$ and $y$ directions, respectively, to form a 2D sequence with the size of $800 \times 800$ by using a logistic XOR gate. An XOR gate performs an exclusive-or operation; that is, a true (1) output results if one, and only one, of the inputs to the gate is true. The output is false (0) if both inputs are false or both are true.

**Section 2. Mathematical proof of the broadband $\pi$-phase difference and the meta-atom design**

To elucidate the mechanism of the broadband $\pi$-phase difference of the TMS meta-atom shown in Fig. 2b, we first studied the reflection phase of the meta-atom by using a two-interface interference model as shown in Supplementary Fig. 1, where the reflecting patch along with the substrate below is modeled as an equivalent surface. The reflection of the meta-atom is then the superposition of the multiple reflections, i.e., $r = a + b + c + \cdots$, where $a = r_{12}$ and $b = t_{12}e^{i\delta}r_{23}e^{i\delta}t_{21}$, $c = t_{12}e^{i\delta}r_{23}e^{i\delta}(r_{21}e^{i\delta}r_{23}e^{i\delta})t_{21}$, $\cdots$ Here, numbers 1 to 3 represent three media as depicted in Supplementary Fig. 1; $r$ and $t$ represent the reflection and transmission coefficients; the first and the second numbers in the subscripts show media at the incident and transmitted sides of the interface, respectively, e.g., $r_{12}$ represents the reflection coefficient on the interface between medium 1 and medium 2 under incidence from medium 1; $\delta = -nd\cos(\theta_2)2\pi c/f$ represents the propagation phase in medium 2, where $d$, $n$, $\theta_2$, $c$, and $f$ are the thickness and the refractive index of medium 2, the refractive angle in medium 2, the light speed in a vacuum, and the frequency of the incidence beam. For simplicity, we consider only the first and second-order terms of multiple reflections and assume a normal incidence, i.e., $\theta_1 = 0$. Then we have

$$r \approx a + b = r_{12} + |t_{12}||r_{23}||t_{21}|e^{i\varphi}, \qquad \text{S2(a)}$$

where $\varphi = 2\delta + \varphi_{t_{12}} + \varphi_{r_{23}} + \varphi_{t_{21}}$ is the phase of $b$. Here, $\varphi_{t_{12}}$, $\varphi_{r_{23}}$ and $\varphi_{t_{21}}$ are separately the phase of $t_{12}$, $r_{23}$ and $t_{21}$. According to Fresnel's law, the phase of $r_{12}$ is $\pi$, and $\varphi_{t_{12}}$ and $\varphi_{t_{21}}$ are 0 when the refractive index of medium 1 is smaller than that of medium 2. Therefore, Eq. S1a is reduced to

$$r = -|a| + |b|e^{i(\varphi_{r_{23}} - 2nd \cdot 2\pi f/c)}, \qquad \text{S2(b)}$$

From Eq. S2b, it is found that the phase of $r$ can be freely tuned by changing $d$. Therefore, it is expected that for two meta-atoms with different $d$, e.g., $d_1$ and $d_2$, their reflection coefficients, i.e., $r_1$ and $r_2$ have $\pi$ phase difference, namely: $r_1$ and $r_2$ are opposite to each other in the complex plane. For instance, we assume that the opposite $r_1$ and $r_2$ are separately positive and negative real numbers at the central frequency $f_0$ as shown in the left panel in Fig. 2c, which requires that the phase of $b$ for the two meta-atoms are separately

$$\varphi_{b_1} = \varphi_{r_{23}} - 2nd_1 \cdot \frac{2\pi f_0}{c} = 0 \qquad \text{S3(a)}$$

and

$$\varphi_{b_2} = \varphi_{r_{23}} - 2nd_2 \cdot \frac{2\pi f_0}{c} = \pi. \qquad \text{S3(b)}$$

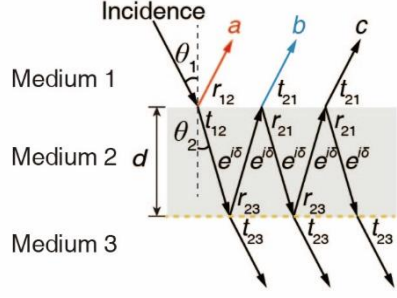

**Supplementary Fig. 1** | Illustration of the interference model of the TMS meta-atom.

In the following, we investigated the conditions for broadband $\pi$ phase difference of $r_1$ and $r_2$ based on the above analysis. It is found in Eq. S2b that when the frequency changes from $f_0$ to $f_0 + \Delta f$, the phase change of the second term of $r_1$ and $r_2$, that is $b_1$ and $b_2$, are separately

$$\alpha = 2nd_1 \cdot 2\pi\Delta f/c \qquad \text{S4(a)}$$

and

$$\beta = 2nd_2 \cdot 2\pi\Delta f/c \qquad \text{S4(b)}$$

, considering that $\varphi_{r_{23}}$ is of very weak dispersion (e.g., $\varphi_{r_{23}} \approx 0.7\pi$ for gold reflecting patches with a thickness ranging from 20 nm to 40 nm). Clearly, $b_1$ and $b_2$ are not opposite to each other anymore at $f_0 + \Delta f$ (solid arrows in the right panel in Fig. 2c) because $\alpha$ and $\beta$ are unequal. Interestingly, under such circumstances, opposite $r_1$ and $r_2$ could still be guaranteed at $f_0 + \Delta f$ as long as $a$ (red arrow in the right panel in Fig. 2c) imposes proper phase corrections $\gamma_1$ and $\gamma_2$ on $r_1$ and $r_2$ to compensate for the phase difference between $\alpha$ and $\beta$, that is $\beta - \gamma_2 = \alpha + \gamma_1$. From the geometric relationship in the right panel in Fig. 2c, it is found that firstly, $\gamma_1$ and $\gamma_2$ can be separately expressed as $\gamma_1 = \arcsin\frac{|a|\sin(\alpha)}{\sqrt{|a|^2+|b|^2-2|a||b|\cos(\alpha)}}$ and $\gamma_2 = \arcsin\frac{|a|\sin(\pi-\beta)}{\sqrt{|a|^2+|b|^2-2|a||b|\cos(\pi-\beta)}}$, and secondly, $\gamma_1 = \gamma_2 = \gamma$ when $r_1$ and $r_2$ are opposite to each other. Therefore, we obtain that the condition leading to the opposite $r_1$ and $r_2$ at $f_0 + \Delta f$ should be

$$\beta = \alpha + 2\gamma = \alpha + 2\arcsin\frac{\sin(\alpha)}{\sqrt{1^2 + (\frac{|b|}{|a|})^2 - 2\frac{|b|}{|a|}\cos(\alpha)}}. \qquad \text{S5}$$

The same condition applies to $f_0 - \Delta f$ when $r_1$ and $r_2$ are located on the real axis of the complex plane at $f_0$ as we assumed above.

Then two meta-atoms with the broadband $\pi$-phase difference in reflection could be designed according to conditions in Eq. S3 and Eq. S5. Among the two equations, Eq. S3 can be easily satisfied by choosing the proper values of $d_1$ and $d_2$. While for Eq. S5, an independent variable to be designed to make the equation hold is $|b|/|a|$, namely, the amplitude ratio of the first-order reflection coefficient to the second-order reflection coefficient within the spectrum of $f_0 \pm \Delta f$.

In the following, we take a practical case of a gold reflecting patch embedded in a silica substrate ($n = 1.5$) as an example. According to Eq. S3, we have $d_1 = 63$ nm and $d_2 = d_1 + 90$ nm at the central frequency of 555 THz. With given $d_1$ and $d_2$, $\alpha$ and $\beta$ are linear functions concerning $\Delta f$ according to Eq. S4. By substituting $\alpha$ and $\beta$ into Eq. S5, the required $|b|/|a|$ for the perfect broadband $\pi$-phase difference within the frequency range of $555 \pm 200$ THz can be obtained. It is found that the required $|b|/|a|$ is very close to 2 with a weak dispersion. We then investigate the tolerance of the value of $|b|/|a|$ for broadband $\pi$-phase difference, considering the difficulty in engineering the dispersion of $|b|/|a|$. The phase difference between $r_1$ and $r_2$, which can be expressed as $\Delta\varphi = \pi + (\alpha + \gamma_1) - (\beta - \gamma_2)$, is calculated for $|b|/|a|$ equals to 1.1, 1.5, 2, 3, 4, and 5 separately and shown in Supplementary Fig. 2. Interestingly, it is found that the phase difference for $|b|/|a| = 2$ approaches to $\pi$ with negligible variations (red solid line) in a broad spectrum of $555 \pm 200$ THz. In addition, the phase difference for $|b|/|a|$ larger than 1.5 and smaller than 3 still approaches $\pi$ with variations less than $0.15\pi$. Therefore, the broadband $\pi$-phase difference can be obtained in the frequency range of $555 \pm 200$ THz as long as $|b|/|a|$ is within a range of around 1.5 to 3. Take into consideration that $|a| = 0.2$ is non-dispersive in this practical case, a broadband $\pi$-phase difference of the two meta-atoms can be obtained when $|b|$ is within the range of 0.3 to 0.6, which can be easily realized by adopting the gold reflecting patch with proper thickness $d_g$, e.g., $d_g = 28$ nm. On the contrary, when the interference mechanism is absent, e.g., $|a| = 0$, the $\pi$-phase difference is limited to $f_0$ as shown in Supplementary Fig. 2 (solid black line).

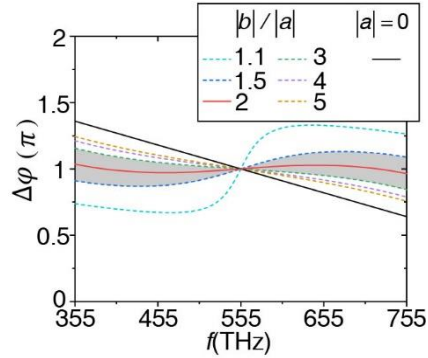

**Supplementary Fig. 2** | Phase difference between $r_1$ and $r_2$ for different values of $|b|/|a|$. For $|b|/|a|$ within a range of 1.5 to 3, the phase difference is very close to $\pi$ in the whole spectrum (the shaded region).

We note that in the above analysis, some simplifications are adopted. The reflection coefficient of realistic meta-atoms would deviate slightly from the theoretical calculation, though such simplifications are reasonable and helpful in revealing the mechanism of the broadband $\pi$-phase difference. Therefore, in designing the practical case of a 28 nm gold reflecting patch in the silica substrate, we fix $d_2 - d_1$ at 90 nm and sweep the value of $d_1$ to calculate the accurate reflection and transmission coefficients spectrum, which are shown in Fig. 2d. It is found that when $d_1$ is around 70 nm, a broadband $\pi$-phase difference is obtained in the whole visible spectrum, which is very close to that in the theoretical analysis, i.e., $d_1 = 63$ nm.

**Supplementary Video Legends**

**Supplementary Video 1: Appearance of the TMS-coated GW (28nm, Au) under outdoor conditions with natural light.** A clear view of the campus is observed through the TMS-coated GW with a matte appearance. This video is shot with an iPhone 12.

**Supplementary Video 2: Augmented reality with the TMS-coated GW (28 nm, Au).** A coffee mug, a tennis ball, and an analog clock are successfully identified and their names are projected on the TMS-coated GW.

**Supplementary Video 3: Augmented reality with a commercial AF.** A coffee mug, a tennis ball, and an analog clock behind the AF can't be identified.